\documentclass[aps,prb,twocolumn,showpacks]{revtex4}
\usepackage{graphicx}  
\usepackage{dcolumn}   
\usepackage{bm}        
\usepackage{amssymb}   
\usepackage{braket}
\usepackage{color}
\usepackage{soul}

\begin{document}

\widetext

\setlength{\textwidth}{175.0mm}


\title{Radiative coupling of A and B excitons in ZnO} 
\date{\today}

\author{Takashi Kinoshita}
\author{Hajime Ishihara}
\affiliation{Department of Physics and Electronics, Osaka Prefecture University, Sakai, Osaka 599-8531, Japan}

\begin{abstract}
Radiation-induced coupling between A and B excitons in ZnO is theoretically studied.
Considering the center-of-mass motion of excitons in bulk and thin film structures, we calculate the eigenmodes of an exciton--radiation coupled system and reveal the ratio of each excitonic component in the respective eigenmodes, which is determined from diagonalization of the self-consistent equation between the polarization and the Maxwell electric field.
In particular, in a nano-to-bulk crossover size regime, the large interaction volume between multipole-type excitonic waves and radiation waves causes radiative coupling between excitons from different valence bands, which leads to an enhancement of the radiative correction.
The results presented in this study are in striking contrast with the conventional view of the optical response of excitons in ZnO, where A and B excitons are independently assigned to their respective spectral structures. 
We demonstrate an alternative spectral assignment of nonlinear optical signals by focusing on the degenerate four-wave mixing.
\end{abstract}

\maketitle


\section{INTRODUCTION}

  ZnO shows great potential for application in optoelectronic devices because of its wide band gap and large exciton binding energy.
In particular, its high excitonic stability has received considerable attention for various applications, including light-emitting diodes \cite{Min,Li}, ultraviolet photovoltaics \cite{Cole}, and exciton-polariton lasing \cite{Orosz,Feng}. 
Furthermore, technologies relevant to the nanofabrication of this material have been rapidly developed \cite{Makino,Ting}, and confined excitons are becoming attractive because of their strengthened excitonic effect.
In terms of the excitonic property, ZnO has a multicomponent nature arising from the nearly-degenerate valence bands \cite{Klingshirn}.
The symmetry ordering of the top valence band in ZnO was first determined by Thomas and Hopfield from the polarization dependences of reflectivity and absorption spectra \cite{Thomas,Hopfield}.
They identified that the topmost valence band has the $\Gamma_7$ symmetry, unlike the other wurtzite compounds.
After that, however, Reynolds {\it et al.} \cite{Reynolds} investigated optically unallowed excitons by the mageto-optical responses, and identified the $\Gamma_9$ symmetry for the topmost valence band as well as the other wurtzite compounds.
Since their work, studies identifying both $\Gamma_7$ \cite{Lambrecht,Rodina,Wagner} and $\Gamma_9$ \cite{Gil,Adachi,Hazu} symmetry have been reported.
From the group theory, the 1s-exciton ground-state symmetry in wurtzite-ZnO is represented as \cite{Honerlage}

\begin{eqnarray}
\Gamma_1\otimes\Gamma_7\otimes\Gamma_9=\Gamma_5\oplus\Gamma_6,
\end{eqnarray} 
for $\Gamma_7$ conduction band and $\Gamma_9$ valence band, while

\begin{eqnarray}
\Gamma_1\otimes\Gamma_7\otimes\Gamma_7=\Gamma_1\oplus\Gamma_2\oplus\Gamma_5,
\end{eqnarray} 
for $\Gamma_7$ conduction band and $\Gamma_7$ valence band.
The excitons related to the transitions from the topmost and the second-most valence bands are labeled A and B excitons, respectively.
Both A and B excitons with $\Gamma_5$-symmetry are dipole-allowed under the condition of $\bm{E}\perp\bm{c}$ and featured in this article.

Because A and B excitons exist in a close energy region, ZnO is expected to form complicated exciton--radiation coupled states.
In recent years, a lot of studies have been performed on ZnO nano- and micro-cavities, in which several excitonic branches strongly couple via cavity photons forming cavity-polaritons \cite{Vugt,Faure,Kawase}. 
Even for the cases without cavities, such coupling between different exciton components via radiation becomes remarkable if the excitonic center-of-mass (CM) wavefunctions maintain coherence in a whole sample, thus forming new eigenmodes consisting of multiple excitonic components.
In a bulk crystal, an eigenstate of the exciton--radiation coupled system is well known as the bulk polariton. 
The dispersion relation of this state in ZnO can be obtained by applying the Pekar's theory \cite{Pekar} to the multiple resonances of excitons. \cite{Skettrup,Lagois1,Hummer}.
Thus, respective peak structures of optical spectra generally contain contributions from the multiple components of excitons.
This classical approach has widely been used to determine the basic excitonic parameters from the linear responses \cite{Lagois3,Lagois2,Cobet}.
However, the coupling effect between different excitons has seldom been noted and interpretations of observed spectra remain ambiguous though the optical properties of A and B excitons in ZnO have been discussed in numerous studies.

The purpose of this paper is to clarify how the radiation-induced coupling between multicomponent excitons appears in the optical spectra by affecting the energy shifts and radiative widths of the coupled eigenmodes. 
The results will provide an alternative way of attributions of the optical signals in ZnO, which is in contrast to the conventional one where the A and B excitons are independently assigned to their respective peak structures.
This would be also important on the controversial problems such as the symmetry ordering of ZnO valence bands because the discussions have been mainly made based on the conventional way \cite{Thomas,Hopfield,Lambrecht,Rodina,Wagner,Reynolds,Gil,Adachi,Hazu}.
The coupling between different excitonic states via radiation should be noted, particularly in a nano-to-bulk crossover size regime. 
In this regime, the coherence length of the CM wavefunction of excitons attains a size on the sub-micron scale, which violates the long-wavelength approximation (LWA).
Accordingly, a wave-wave coupling between excitons and radiation occurs, and their interaction volume becomes considerably larger, especially for multipole-type excitons with the CM quantum number $\lambda\geq2$, leading to large radiative corrections. 
As a result, the nonlinear optical responses are enhanced \cite{Ishihara3,Amakata,Ishihara}, and the level shift and radiative decay rate are resonantly increased \cite{Syouji,Kishimoto,Ichimiya1} with the sample size.
We should note that such a large interaction volume also strengthens the coupling between different components of excitons via radiation.
From this viewpoint, it is interesting to examine, particularly beyond the LWA regime, 
how the radiation-induced coupling between A and B excitons modifies the optical spectra relative to those expected only from the single-component excitonic systems.

In the present paper, we theoretically demonstrate the exciton--radiation coupled modes and the optical responses of ZnO in two cases, i. e., a semi-infinite system and thin film structures. 
In these demonstrations, we explicitly treat the spatial structures of both the radiation field and the excitonic CM wavefunctions to fully consider their self-consistency affecting the optical responses of multicomponent excitons. 
The results clearly show that the radiation-induced coupling of A and B excitons plays an essential role in the formation of exciton--radiation coupled modes in both cases.
We successfully clarify a component ratio of A and B excitons in the exciton--radiation coupled modes, which shows a strong band-mixing of excitons owing to the radiative coupling.
In particular, this effect leads to an enhancement of the level shifts and radiative widths in thin film geometry 
that is directly reflected in the linear and nonlinear optical spectra.
The results would change the simple interpretation of optical signals of multicomponent excitons where each component is independently assigned to their respective peak structures.

  The rest of this article is organized as follows:
Section \ref{2} outlines the theory of nonlocal optical response considering a self-consistent interplay between multicomponent excitons and radiation.
Section \ref{3} describes how the A and B excitons couple via radiation and appear in the reflectivity spectra in a semi-infinite system.
In Sec. \ref{4}, we clarify the relation between anomalous exciton--radiation coupled modes in thin film structures and their nonlinear optical responses by considering DFWM signals as an example.
The results and discussions in this article are summarized in Sec. \ref{5}.  

\section{Nonlocal optical response of multicomponent excitons\label{2}}
  We consider a sample with the film thickness much larger than the excitonic Bohr radius and that is periodic along the film surface.
In this condition, the relative motion of an exciton can be treated in the same way as those in a bulk, although the CM motion is confined in a thickness direction.
According to the standard effective-mass approximation, the eigenenergy of the unperturbed excitonic system is written as

\begin{eqnarray}
E_{\sigma\lambda}=E_\sigma+\frac{\hbar^2k^2_{\sigma\lambda}}{2M_\sigma},
\end{eqnarray}
where $\sigma$ is an index to label multiple exciton bands, $\lambda$ is an index of quantized excitonic state, $E_\sigma$ is the transverse energy of exciton at bulk limit, $k_{\sigma\lambda}$ is a wavenumber satisfying the quantization condition, and $M_\sigma$ is the effective mass of exciton.
From the translational symmetry along the surface direction, an excitonic wavefunction of the CM motion is given as
$
\psi_{\sigma\lambda}(\bm{r})=g_{\sigma\lambda}(z)S^{-\frac{1}{2}}e^{i\bm{k}_{\parallel}\cdot\bm{r}_{\parallel}},
$
where $g_{\sigma\lambda}(z)$ is the CM wavefunction in the thickness direction, $S$ is a unit area along the film surface, and $\bm{k}_{\parallel}$ and $\bm{r}_{\parallel}$ are the lateral components of the wavevector and position vector, respectively.

  To describe the self-consistent interplay between the spatial structures of the radiation field and excitonic waves, we apply the nonlocal response theory \cite{Cho2} to the multicomponent excitonic system.
The standard expression of the exciton--radiation interaction Hamiltonian is expressed as

\begin{eqnarray}
{\cal H}_{int}=-\int\hat{\bm{{\cal P}}}(\bm{r})\cdot\tilde{\bm{{\cal E}}}(\bm{r},t)d\bm{r},
\label{H_int}
\end{eqnarray}
where $\hat{\bm{{\cal P}}}(\bm{r})$ is the polarization operator for electrons integrated over the cell at $\bm{r}$ \cite{Cho2}, and $\tilde{\bm{{\cal E}}}(\bm{r},t)$ is the Maxwell electric field. 
This interaction depends not only on their amplitudes but also spatial structures and interaction volume, which exhibit the nonlocal effect.
The matrix element of the polarization operator can be written as \cite{Iida}
$
\bra{0}\hat{\bm{{\cal P}}}(\bm{r})\ket{\psi_{\sigma\lambda}}=\mu_\sigma\psi_{\sigma\lambda}(\bm{r})\,\vec{e}_p,
$
where $\vec{e}_p$ is a unit vector in the polarization direction.
Note that, in our definition, $\mu_\sigma$ has the dimension of dipole moment per one-half power of volume.
This value is determined from the multiple Longitudinal-Transverse (LT) splitting energies as shown in the third paragraph in Sec. \ref{3}.

  According to the density matrix method \cite{Shen}, the $j$-th order polarization field can be obtained from $\tilde{\bm{{\cal P}}}^{(j)}(\bm{r},t)={\rm Tr}[\rho^{(j)}(t)\hat{\bm{{\cal P}}}(\bm{r})]$, where $\rho^{(j)}(t)$ is the $j$-th order density matrix.
Assuming the electric field to be $\bm{{\cal E}}(\bm{r},\omega)={\cal E}(z,\omega)S^{-\frac{1}{2}}e^{i\bm{q}_\parallel\cdot\bm{r_\parallel}}\,\vec{e}_p$, where $\bm{q}_{\parallel}$ is the lateral component of the wavevector $\bm{q}$ of light in vacuum,
then the first-order polarization is expressed in one-dimensional form as \cite{Cho2,Ishihara2}

\begin{eqnarray}
{\cal P}^{(1)}(z,\omega)
&=&
\int\chi(z,z',\omega){\cal E}(z',\omega)dz'.
\label{nonlocal}
\end{eqnarray}
In this expression, a resonant term of the nonlocal susceptibility is written as 

\begin{eqnarray}
\chi(z,z',\omega)
=
\sum_{\sigma}\sum_{\lambda}\frac{p_{\sigma\lambda}(z)p^\ast_{\sigma\lambda}(z')}{E_{\sigma\lambda}-\hbar\omega-i\Gamma_\sigma},
\label{chi}
\end{eqnarray}
where $\Gamma_\sigma$ is a nonradiative damping constant and $p_{\sigma\lambda}(z)=\mu_\sigma g_{\sigma\lambda}(z)$.

  Assuming the normal incidence for simplicity, the polarization field from the resonant contribution should be determined self-consistently with the following Maxwell equation \cite{Cho2,Ishihara2}:

\begin{eqnarray}
(-\frac{\partial^2}{\partial z^2}-\epsilon_{b}q^2){\cal E}(z,\omega)
=4\pi q^2{\cal P}(z,\omega),
\label{1dimension}
\end{eqnarray}  
where $q=|\bm{q}|$ and the background dielectric constant $\epsilon_b$ indicates the contribution from the nonresonant polarization.
The solution of Eq. (\ref{1dimension}) can be described with a retarded Green's function\cite{Chew} that satisfies

\begin{eqnarray}
(-\frac{\partial^2}{\partial z^2}-\epsilon_{b}q^2){\cal G}(z,z',\omega)
=\delta(z-z'),
\end{eqnarray} 
where $\delta(z-z')$ is the delta function.
Using ${\cal G}(z,z',\omega)$, the Maxwell electric field can be written in integral form as

\begin{eqnarray}
{\cal E}(z,\omega)
&=&
{\cal E}^{(0)}(z,\omega)
+
4\pi q^2\int {\cal G}(z,z',\omega){\cal P}(z',\omega)\,dz',
\nonumber
\\
\label{maxwell}
\end{eqnarray}
where ${\cal E}^{(0)}(z,\omega)$ is the background electric field.
By considering Eqs. (\ref{nonlocal}), (\ref{chi}), and (\ref{maxwell}) in the framework of the linear response where $P(z,\omega)=P^{(1)}(z,\omega)$, the Maxwell electric field ${\cal E}(z,\omega)$ can be rewritten  as

\begin{eqnarray}
&&\hspace{-10mm}
{\cal E}(z,\omega)
=
{\cal E}^{(0)}(z,\omega)
\nonumber
\\
&&\hspace{-5mm}+\,
4\pi q^2\sum_{\sigma}\sum_{\lambda}\int{\cal G}(z,z',\omega)p_{\sigma\lambda}(z')\,dz'X_{\sigma\lambda}(\omega),
\end{eqnarray}
where

\begin{eqnarray}
X_{\sigma\lambda}(\omega)
=
\frac{1}{E_{\sigma\lambda}-\hbar\omega-i\Gamma_\sigma}\int \,p^\ast_{\sigma\lambda}(z){\cal E}(z,\omega)\,dz
\end{eqnarray}
indicates an amplitude of the polarization related to the $\lambda$-th $\sigma$-band exciton component.
Then, we can obtain a closed linear equation system to determine $X_{\sigma\lambda}$ as \cite{Kishimoto}

\begin{eqnarray}
(E_{\sigma'\lambda'}-\hbar\omega-i\Gamma_{\sigma'})X_{\sigma'\lambda'}&
\nonumber
\\
&\hspace{-15mm}
+\sum_{\sigma}\sum_{\lambda}Z_{\sigma'\sigma\lambda'\lambda}X_{\sigma\lambda}=F^{(0)}_{\sigma'\lambda'},
\label{self-consistent}
\end{eqnarray}
where $F^{(0)}_{\sigma'\lambda'}(\omega)=\int p^\ast_{\sigma'\lambda'}(z){\cal E}^{(0)}(z,\omega)\,dz$ means an interaction between the exciton and the background electric field.
In addition, $Z_{\sigma'\sigma\lambda'\lambda}$ is the radiative correction from the bare exciton energy written as

\begin{eqnarray}
Z_{\sigma'\sigma\lambda'\lambda}
=-4\pi q^2\int\int 
p^\ast_{\sigma'\lambda'}(z)
{\cal G}(z,z',\omega)
p_{\sigma\lambda}(z')\,dzdz',
\nonumber
\\
\label{correction}
\end{eqnarray}
which indicates the coupling between $\lambda$-th $\sigma$-band exciton and $\lambda'$-th $\sigma'$-band exciton via radiation.
This term also includes a radiation-induced coupling between different band excitons (A and B excitons for ZnO) when $\sigma'\neq\sigma$.
By describing Eq. (\ref{self-consistent}) in a matrix form as $\bm{S}\bm{X}=\bm{F}^{(0)}$, the roots of
$
{\rm det}|\bm{S}|=0
\label{det}
$
provide the eigenmodes of the exciton--radiation coupled system\cite{Kishimoto} as demonstrated in Secs. \ref{3} and \ref{4}.
Also, the density of each excitonic component $D_{\sigma\lambda}$ can be defined from the eigenfunction $X_{\sigma\lambda}$ as

\begin{eqnarray} 
D_{\sigma\lambda}=\frac{|X_{\sigma\lambda}|^2}{\sum_{\sigma'\lambda'}|X_{\sigma'\lambda'}|^2}.
\end{eqnarray}
Using this quantity, we can discuss the ratio of each excitonic component in the exciton--radiation coupled modes. 

The present approach can be applicable to various situations where the quantized multicomponent excitons interact with the radiation field. 
In this paper, we choose appropriate parameters to demonstrate the radiative coupling of A and B excitons in ZnO.

\section{Semi-infinite system \label{3}}
  Before we examine the film geometry, it is interesting to note how much the A and B excitonic components are mixed in the three exciton-polariton branches, called the upper polariton branch (UPB), middle polariton branch (MPB), and lower polariton branch (LPB) in a semi-infinite system.
The exciton--polariton dispersion and reflectivity spectra have been calculated considering the multicomponent excitons and utilizing the additional boundary condition methods \cite{Hummer,Lagois1,Lagois2}.
However, thus far, they have not been discussed from the view point of a ratio of each excitonic component.

In a semi-infinite system where the film thickness $d\rightarrow\infty$, we neglect the distortion of CM wavefunctions near the surface and assume $g_{\sigma\lambda}(z)=(2/d)^{1/2}\sin(k_\lambda z)$ in which $k_\lambda$ satisfies the quantization condition: $k_{\lambda}d=\lambda\pi$ $(\lambda=1,2,\cdots)$.
By utilizing the Green's function ${\cal G}(z,z',\omega)$ for a free space\cite{Chew}, the polariton dispersion relation can be obtained from
$
{\rm det}|\bm{S}|=0
$
as

\begin{eqnarray}
\frac{c^2k^2_\lambda}{\omega^2}=\epsilon_b+\sum_{\sigma=A,B}\frac{4\pi|\mu_\sigma|^2}{E_{\sigma\lambda}-\hbar\omega-i\Gamma_\sigma}
\equiv
\epsilon_p(\omega),
\label{polariton_despersion}
\end{eqnarray}
where $\epsilon_b=3.7$ for ZnO \cite{Madelung}.
By introducing wavenumbers $k_i$ (${\rm Re}[k_i]\geq0$, ${\rm Im}[k_i]\geq0$, $i$=1,2,3) for the three polaritons, the Maxwell electric field can be written as

\begin{eqnarray} 
{\cal E}(z,\omega)
=
\sum_{i}\zeta_i(\omega)\{{\cal E}_1\sin{k_iz}+{\cal E}_2\sin{k_i(d-z)}\}, 
\end{eqnarray}
where ${\cal E}_{1}$ (${\cal E}_{2}$) is an arbitrary constant, and $\zeta_i(\omega)$ indicates an amplitude of each polariton component.
Here, we assumed the background has a small absorption , i. e., $e^{iQd}\approx0$ ($d\rightarrow\infty$).
The Maxwell's boundary conditions of electromagnetic field at the incident surface provide the reflectivity spectrum.
The detailed calculation is shown in the appendix.

In eq. (\ref{polariton_despersion}), $|\mu_\sigma|^2$ is related to the two longitudinal exciton energies $E_{L1}$ and $E_{L2}$ ($E_{L1}<E_{L2}$) \cite{Lagois1} which are obtained as the roots of $\epsilon_p(\omega)=0$ at $k_\lambda=0$ and $\Gamma_\sigma=0$.
By using these quantities, $|\mu_\sigma|^2$ can be rewritten as
$
|\mu_A|^2=\epsilon_b\frac{\Delta_{LT1}}{4\pi}\frac{E_{L2}-E_A}{E_B-E_A}
$
and 
$
|\mu_B|^2=\epsilon_b\frac{\Delta_{LT2}}{4\pi}\frac{E_B-E_{L1}}{E_B-E_A},
$
where $\Delta_{LT1}$ $(=E_{L1}-E_A)$ and $\Delta_{LT2}$ $(=E_{L2}-E_B)$ is the LT splitting energies.
Table \ref{parameters} lists the parameters of bulk ZnO \cite{Lagois2}, where $m_0$ is the static electron mass.
From these parameters, we can obtain $|\mu_A|^2\approx1.570\times10^{-3}$ eV and $|\mu_B|^2\approx1.964\times10^{-3}$ eV.

\begin{table}[h]
\caption{Parameters of bulk ZnO \cite{Lagois2}}
\begin{tabular}{l@{\hspace{25mm}}l}
\hline \hline
\hspace{10mm}A & \hspace{10mm}B \\ \hline
$M_A=0.87m_0$ & $M_B=0.87m_0$ \\
$E_A=3.3758$ eV & $E_B=3.3810$ eV \\
$E_{L1}=3.3776$ eV & $E_{L2}=3.3912$ eV \\
$\Delta_{LT1}=1.8$ meV & $\Delta_{LT2}=10.2$ meV \\
\hline
\label{parameters}
\end{tabular}
\end{table}

\begin{figure}[h]
\begin{center}
\includegraphics[width=85mm]{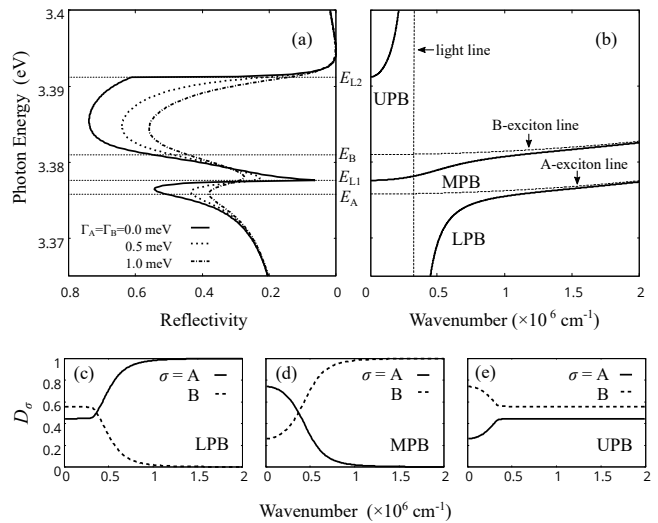}
\caption{(a) Calculated reflectivity of a semi-infinite ZnO in the A and B exciton energy region for three nonradiative damping ($\Gamma_\sigma$) values,\hspace{1.5mm}(b) Polariton dispersion relation, and the diagonalized density of A and B excitonic components $D_\sigma$ in (c) LPB (d) MPB, and (e) UPB.
The horizontal dashed lines in Fig. (a) indicate the energy positions at $E_A$, $E_{L1}$, $E_B$, and $E_{L2}$, in order of increasing energy.
The light line and the $\sigma$-exciton line in Fig. (b) indicate $\hbar\omega=\hbar ck/\sqrt{\epsilon_b}$ and $\hbar\omega=E_{\sigma}+(\hbar^2k^2)/(2M_\sigma)$, respectively.}
\label{reflectivity}
\end{center}
\end{figure}
Figure \ref{reflectivity}(a) and (b) show the calculated reflectivity of a semi-infinite ZnO, and the polariton dispersion relationship, respectively.
The characteristic point is a large difference in the two LT splitting energies.
This difference does not result from a change in the polarizability or the oscillator strength, but merely from the interaction between the two resonances \cite{Lagois1}.
In other words, the LT splitting depends not only on the oscillator strength but also the energy separation between different exciton resonances.

  The interaction between different exciton branches also causes the component-mixing in polariton states.
Figures \ref{reflectivity}(c), (d), and (e) show the diagonalized density of each exciton component $D_\sigma$ in LPB, MPB, and UPB, respectively.
If the energy separation $E_A-E_B$ is much larger than the LT splitting energies, the changes in each polariton nature are simple with increase in the wavenumber as follows: 
The LPB changes from the A-B mixed photon-like state to the A exciton-like state.
The MPB changes from the A exciton-like state to the B exciton-like state through the A-B mixed photon-like state.
The UPB changes from the B exciton-like state to the A-B mixed photon-like state.
However, with appropriate ZnO parameters in Table \ref{parameters} where $E_B-E_A$ is comparable to the LT splitting energies, the MPB and the UPB include both A and B excitonic components not only in the photon-like region but also in the zero-wavenumber region as shown in Figs. \ref{reflectivity}(d) and (e). 
These results indicate, even in a semi-infinite system, the A and B excitons should not be assigned independently to the optical peak signals, for example, to the dip structure at $E_{L1}$ in Fig. \ref{reflectivity}(a) which can be understood from the MPB's propagation in the sample.

\section{Thin film \label{4}}
  The coupling of the A and B excitons shows more exotic behavior in the confined geometry.
Although there have been no experimental reports on the quantization of the excitonic CM motions for ZnO, it can be observed as the other materials\cite{Tang,Nakayama} if the samples have sufficient quality with small nonradiative damping constants.
In a thin sample, the distortion of wavefunctions near the surface generally affects the energy structures of exciton.
We therefore applied microscopic transition layer (TL) model \cite{DelSole,Ishihara2} as the CM wavefunctions of exciton.
In this model, the quantization condition is given as 
$
k_{\sigma\lambda}d-2\tan^{-1}{k_{\sigma\lambda}/P_\sigma}=\lambda\pi
$
($\lambda=$ $1,2,\cdots$), where $P_\sigma$ is a decay constant of evanescent waves with a value on the order of the inverse of the effective Bohr radius indicating the distortion length.
This model provides proper excitonic level structures even when the distortion cannot be regarded as being negligibly small in comparison with the sample thickness, 
Furthermore, it has been discussed that the distortion contributes not only to the excitonic energies but also to the spectral shapes because the shape of the wavefunctions changes with the thickness under a particular value of $P_\sigma$ \cite{Yoshimoto}.
This is why the model can be utilized to determine some excitonic parameters accurately by reproducing both spectral shapes and quantized CM levels of excitons observed in experiments as demonstrated in Ref. \cite{Yoshimoto}.
In this paper, however, we fix these values as the effective Bohr radius\cite{Klingshirn} ($1/P_A=1/P_B=1.8$ nm) because $P_\sigma$ does not affect the essence of the radiative coupling of A and B excitons although it would be a powerful tool for accurate analyses of the CM quantization of excitons even in the LWA regime.

  Now, we discuss the exciton--radiation coupled states in the thin film beyond the LWA regime.
By utilizing the Green's function ${\cal G}(z,z',\omega)$ for a slab structure\cite{Chew}, the complex eigenmodes $\hbar\omega_{\xi}$ of the exciton--radiation coupled system can be obtained from
$
{\rm det}|\bm{S}|=0,
$
where $\xi$ is an index of quantized coupled states.
The real part Re[$\hbar\omega_{\xi}$] gives the eigenenergy including the radiative shift and the imaginary part $-$Im[$\hbar\omega_{\xi}$] gives the radiative width.
An increase in the film thickness leads to a large interaction volume between excitonic waves and radiation waves, and their phase-matching leads to the large radiative correction from the bare excitonic states especially in a nano-to-bulk crossover size regime \cite{Amakata,Ishihara,Syouji,Kishimoto,Ichimiya1}.

\begin{figure}[h]
\begin{center}
\includegraphics[width=85mm]{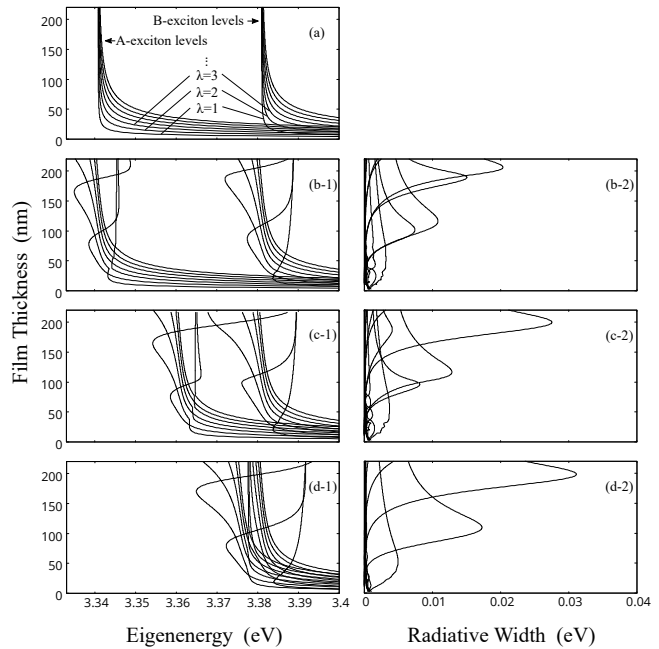}
\caption{Dependences of eigenenergies and radiative widths on $E_B-E_A$, which is virtually varied to demonstrate the radiative coupling between the A and B excitons. $E_B-E_A=$ (a) 40.2 meV with the velocity of light $c$ assumed to be infinite,\hspace{1.5mm}(b-1,2) 40.2 meV,\hspace{1.5mm}(c-1,2)  20.2meV, and\hspace{1.5mm}(d-1,2) 5.2 meV which is the same parameter as listed in Table \ref{parameters}.}
\label{eigenmode}
\end{center}
\end{figure}

To see how the A and B excitons contribute to the coupled mode scheme in thin films, we examine several hypothetical values of $E_B-E_A$ (Fig. \ref{eigenmode}). 
In this calculation, we fix the intensities of the transition dipole density to $|\mu_A|^2\approx1.570\times10^{-3}$ eV and $|\mu_B|^2\approx1.964\times10^{-3}$ eV. Figure \ref{eigenmode}(a) shows the bare excitonic modes, namely, the case where the retarded interaction providing the intrinsic radiative width of exciton is absent. 
Thus, the eigenenergies reach $\hbar\omega\approx E_\sigma+\hbar^2k^2_{\sigma\lambda}/(2M_\sigma)$, and the radiative widths are zero for both A and B excitons.
It should be noted that we neglect the confinement effect of the relative motions of electron-hole pair, which dominantly contributes to the excitonic energy structures in the size region where the thickness reaches the effective Bohr radius \cite{Kayanuma} (about $1.8$ nm for ZnO).  
Even in the presence of the retarded interaction, when the bare A and B excitons are energetically separated over 40 meV as shown in Fig. \ref{eigenmode}(b), they independently couple with the radiation and form their respective eigenmodes as the single-component excitons.
In this case, a spatial phase-matching between a particular CM wavefunction and the radiation enhances the radiative correction of multipole-type excitons \cite{Kishimoto}.
However, as $E_A$ approaches $E_B$, as shown in Figs. \ref{eigenmode}(c) and (d), the radiation-induced coupling of A and B excitons becomes obvious.
In particular, the components with the same $\lambda$ strongly interact with each other.
Accordingly, either branch dominates the radiative corrections leading to an increase of the energy shift and the radiative width; conversely, the other branch decreases them.
In Fig. \ref{eigenmode}(d), we use the same parameters as those listed in Table \ref{parameters}.
Comparing Figs. \ref{eigenmode}(b-2) and (d-2), we find that the local maximal values of the radiative widths increase owing to the radiation domination by particular coupled modes, which means that the radiative decays of these modes become faster than those considering only the single-component exciton. 

  Such behavior of the thickness-dependent eigenmodes of exciton--radiation coupled system would be expected to affect the peak energy structures and spectral widths in the nonlinear optical signals.
In particular, the analysis of the ratio of excitonic component provides a clear interpretation of nonlinear optical spectra of ZnO from the view point of the radiative coupling between A and B excitons.

\subsection*{DFWM Signals}
In this section, we demonstrate nonlinear optical responses focusing on the degenerate four-wave mixing (DFWM), which is a typical third-order nonlinear process.
The signal light is emitted to the direction $2\bm{k}_2-\bm{k}_1$ in which $\bm{k}_1$ and $\bm{k}_2$ is a wavevector of incident lights.
For simplicity, we assume that the incident lights propagate in the same direction perpendicular to the film surface.
Considering the third-order nonlinearity, we examined the state filling due to the Pauli
exclusion effect and the exciton-exciton interaction.
By discretizing the medium and assuming the one-dimensional
transfer reduced from the effective mass $M_\sigma$,
we introduced an attractive interaction between the excitons at
neighboring sites, which yields biexciton and free two-exciton
states \cite{Ishihara3,Amakata}. 
In the following discussions, the contribution of the biexciton resonance
is not essential because of its large binding energy (although the biexciton signals may experimentally appear in the one-exciton energy region). 
Also, the induced absorption due to the transition from the one-exciton to the free-two exciton state considerably decreases with increase in the size of excitonic system as demonstrated in Ref. \cite{oka}.
The contribution of the free-two exciton states can be estimated in the order of $10^{-3}$ or less in the beyond-LWA regime where the radiative decay of the one-exciton state is enhanced with the thickness.
Thus, in the present demonstration, we focus on the dominant contribution, i.e., the effects of the one-exciton resonance while avoiding non-essential issues of two-exciton contributions.
It should be noted that the elaborate analysis considering the free two-exciton states through the cancellation effect \cite{Ishihara4} is necessary for evaluating the absolute values of DFWM signal.

In the configuration of DFWM, the third-order polarization considering three-fold resonant and multicomponent terms can be described as

\begin{eqnarray}
{\cal P}^{(3)}(z,\omega)
=\sum_{\sigma}\sum_{\nu}
U_{\sigma\nu}(\omega)
p_{\sigma\nu}(z),
\end{eqnarray}
where

\begin{eqnarray}
&
&
\hspace{-5mm}
U_{\sigma\nu}(\omega)
=
\sum_\lambda\int\int d\omega_1d\omega_2\bar{X}_{\sigma\nu\lambda}(\omega, \omega_1, \omega_2) \nonumber 
\\
&&
\times H^{k_2}_{\sigma\nu}(\omega_1)
H^{\ast k_1}_{\sigma\lambda}((\omega_1+\omega_2)-\omega)
H^{k_2}_{\sigma\lambda}(\omega_2).
\label{U}
\end{eqnarray}
In this expression, $H^{k_1(k_2)}_{\sigma\nu}(\omega)=\int p^\ast_{\sigma\nu}(z){\cal E}(z,\omega)\,dz$ should be determined self-consistently by solving the third-order Maxwell equation. 
If, however, we assume the incident intensity regime where the electric field originated from the third-order polarization is much weaker than that originated from the linear polarization, then it is a good approximation that the $H^{k_1(k_2)}_{\sigma\nu}(\omega)$ corresponds to the value obtained from the linear response calculation.
In Eq. (\ref{U}), $\bar{X}_{\sigma\nu\lambda}(\omega, \omega_1, \omega_2)$ includes 
energy denominators of the triple-resonance of $\omega_1$, $\omega_2$ and $\omega$ written as

\begin{eqnarray}
\bar{X}_{\sigma\nu\lambda}(\omega,\omega_1,\omega_2) 
&&
=
\frac{1}
{(\hbar\omega_1-\hbar\omega-i\gamma_\sigma)
(E_{\sigma\nu}-\hbar\omega-i\Gamma_\sigma)}\cdot 
\nonumber
\\
&&\hspace{-30mm}
\{\frac{1}{E_{\sigma\lambda}-\hbar\omega_2-i\Gamma_\sigma}+\frac{1}
{-E_{\sigma\lambda}+\hbar(\omega_1+\omega_2-\omega)-i\Gamma_\sigma}\} \nonumber 
\\
&&
\hspace{-30mm}
+\frac{1}
{(E_{\sigma\nu}-E_{\sigma\lambda}-\hbar\omega+\hbar\omega_2-i\Gamma_\sigma)
(E_{\sigma\nu}-\hbar\omega-i\Gamma_\sigma)} \cdot
\nonumber
\\
&&\hspace{-30mm}
\{
\frac{1}{-E_{\sigma\lambda}+\hbar(\omega_1+\omega_2-\omega)-i\Gamma_\sigma}
+\frac{1}{E_{\sigma\nu}-\hbar\omega_1-i\Gamma_\sigma}
\},
\label{X3}
\end{eqnarray}
where $\gamma_\sigma$ is a nonradiative population decay constant.
Considering the first- and third-order polarizations (namely, ${\cal P}(z,\omega)={\cal P}^{(1)}(z,\omega)+{\cal P}^{(3)}(z,\omega)$), we can write the total electric field ${\cal E}(z,\omega)$ of this configuration as

\begin{eqnarray}
{\cal E}(z,\omega)
&=&
{\cal E}^{(0)}(z,\omega)
\nonumber
\\
&&
\hspace{-10mm}
+
\sum_{\sigma}\sum_{\nu}C_{\sigma\nu}(z,\omega)\{X_{\sigma\nu}(\omega)+U_{\sigma\nu}(\omega)\},
\label{nonlinear}
\end{eqnarray}
where

\begin{equation}
C_{\sigma\nu}(z,\omega)=4\pi q^2\int{\cal G}(z,z',\omega)p_{\sigma\nu}(z')\,dz'.
\end{equation}
Here, we should note that the signals contain pure nonlinear components without the background electric field, i. e., ${\cal E}^{(0)}(z,\omega)=0$.

In previous researches, our theoretical scheme has successfully reconstructed experimental data for single-component excitonic systems such as CuCl \cite{Syouji,Ichimiya1}.
In the case of multicomponent excitonic system, however, the effects of radiative coupling between different excitonic branches would be expected to reflect in the nonlinear optical signals.

\begin{figure}[h]
\begin{center}
\includegraphics[width=85mm]{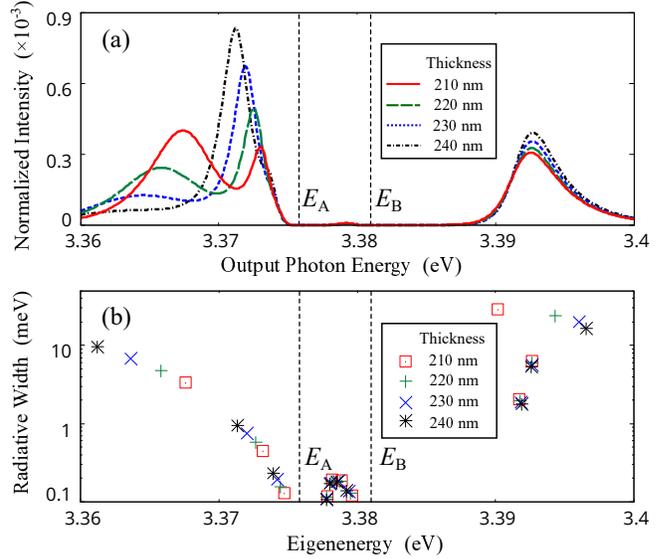}
\caption{(a) Film thickness dependence of the calculated DFWM spectra of a ZnO thin film in the excitonic resonance region normalized by a peak intensity of the incident light. (b) Eigenenegy vs. radiative width of the exciton--radiation coupled modes for the corresponding film thickness. The vertical lines indicate the energies of transverse A and B excitons.}
\label{spectra}
\end{center}
\end{figure}

\begin{figure*}[htbp]
\begin{center}
\includegraphics[width=140mm]{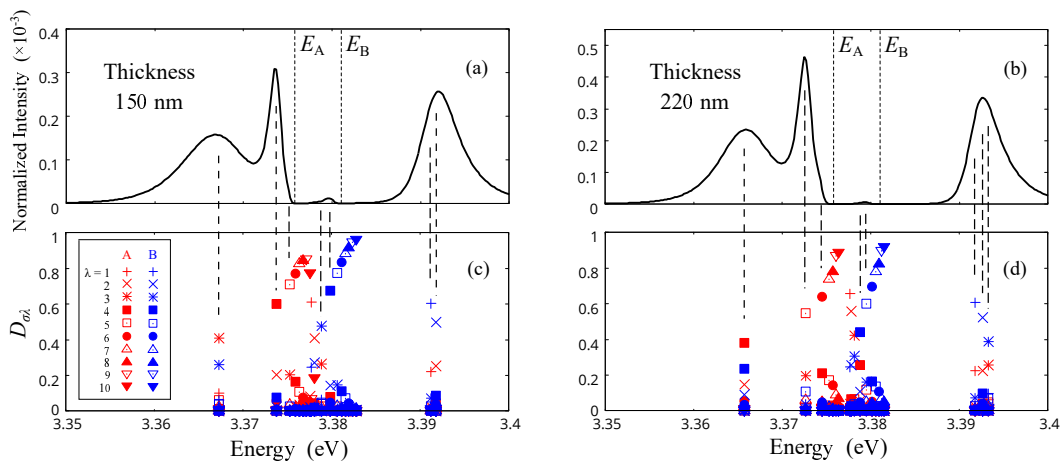}
\caption{DFWM spectra of a ZnO thin film with thickness (a) 150 nm, (b) 220 nm, and
density of each excitonic component $D_{\sigma\lambda}$ with thickness (c) 150 nm, and (d) 220 nm.
The vertical dashed lines across the upper and lower figure indicate the energy positions of the exciton--radiation coupled modes that dominantly appear in the spectra.}
\label{ratio}
\end{center}
\end{figure*}

  As conditions of incident lights, we assume the Gaussian pulses for which the FWHM is 120 fs ($\approx15.2$ meV) to cover the wide spectral region, the integrated intensity is 3.0 $\mu$J/cm$^2$, and the center energy is 3.378 eV.
To clearly show peak structures of the signals, we set the non-radiative damping parameters as $\Gamma_\sigma=\gamma_\sigma=0.5$ meV.
The generated nonlinear signal includes every combination of Fourier components in the pump and probe light.
We integrate over these components by the numerical method.

  Figure \ref{spectra}(a) shows film thickness dependence of the calculated DFWM spectra of a ZnO thin film in the excitonic resonance region normalized by a peak intensity of the incident light.
The peak energies and spectral widths clearly reflect the eigenmodes of exciton--radiation coupled system, as shown in Fig. \ref{spectra}(b).
The radiative widths of the lower (Re[$\hbar\omega_\xi$]$<E_A$) and the upper ($E_B<$Re[$\hbar\omega_\xi$]) branches are broader than those of the middle branch ($E_A\leq$ Re[$\hbar\omega_\xi$]$\leq E_B$) owing to the radiation domination by particular coupled modes as seen in the Figs. \ref{eigenmode}(d-1) and (d-2).
This is why the upper and the lower signals are dominant in the spectra compared with the middle signals.
Accordingly, the splitting with a value larger than $E_{L2}-E_{A}$ between the upper and lower peaks is obvious in the spectra.
In addition, spectral changes with the thickness can also be explained by the thickness-dependent behavior of the exciton--radiation coupled modes.
Particularly in the lower branch, the eigenenergies are red-shifted and the radiative widths are broadened with the thickness from $210$ nm to $240$ nm, which cause the low energy shift and the broadening of the lower DFWM signals, respectively.

  These upper and lower two peaks of the DFWM signal have been experimentally reported for a 55-nm-thick ZnO thin film \cite{Zhang}, which indicates our calculation reproduces an essential profile of observed spectra.  
However, the energy positions of the two peaks look a little different from our calculation results.
In particular, the experimental result seems to include contributions from higher energy components above 3.4 eV, where the continuum electron--hole states not included in the present model might be one of the reasons.

Then, we investigate which component of the A and B excitons is dominant in the peak structure of DFWM signals.
Figure \ref{ratio} shows DFWM spectra of a ZnO thin film with thickness (a) 150 nm and (b) 220 nm, and the density of each excitonic component $D_{\sigma\lambda}$ with thickness (c) 150 nm and (d) 220 nm, which clearly indicates attributions of the DFWM signals.

The radiative coupling between excitons from different valence-bands affects the ratio of excitonic components for the exciton--radiation coupled modes.
There are two important points in this figure.
First, attributions of DFWM peak structures are quite different even though the spectral shapes are similar, as shown in Figs. \ref{ratio} (a) and (b).
For example, the lowest peak with thickness 150 nm is dominantly attributed to the $\lambda=3$ exciton of both A and B excitons, although that with thickness 220 nm is dominantly attributed to the $\lambda=4$ exciton.
This is because the large radiative corrections cause interchanges of the quantized excitonic states \cite{Syouji}.
Second, the mixing of A and B excitons becomes significant with increase in the energy shifts from the bare exciton energies. 
In particular, the lowest and uppermost modes include both A and B exciton components comparably because each component with the same $\lambda$ strongly interact with each other, as shown in Figs. \ref{eigenmode}(d-1) and (d-2).
The energy dependence of $D_{\sigma\lambda}$ is consistent with the case of the bulk system shown in Figs. \ref{reflectivity} (c)-(e) where the mixing of A and B excitons is prominent with the energy shifts in the LPB and UPB.

In most cases, the peak signals of excitons in ZnO are independently assigned to either A or B exciton (for example, in Ref. \cite{Hazu}) without the view point of the radiation-induced component-mixing.
On the other hand, the present results indicate that the coupled modes contributing the signals include both A and B exciton components comparably, which would change the conventional interpretation of the observed spectra.
On the other hand, the observation of the radiative coupling of A and B excitons for the respective CM modes is a challenge because the signature of quantized CM motion of excitons has not been found for ZnO thin films in the past experiments. 
The CM quantization in nano-to-bulk crossover regime will be observed if the larger coherent volume of excitons is realized in ZnO samples with improved crystal quality.

\vspace{3mm}
\section{Conclusion \label{5}}
  In the present work, we have theoretically revealed the crucial role of the radiation-mediated coupling between multicomponent excitons in their optical responses.
For example, in a semi-infinite medium, three polariton branches (upper, middle, and lower branches) include both A and B excitonic components not only in the photon-like region but also in the region around $k=0$.
This understanding provides high transparency to the spectral formation in the exciton resonance region for ZnO.
Furthermore, the radiative coupling exhibits the peculiar thickness-dependent mode structures of the exciton--radiation coupled system in thin films.
The radiation domination by a particular branch leads to an enhancement of the radiative shifts and widths compared with the single-component excitonic system. 
We have also found that the DFWM spectra directly reflect the eigenenergies and radiative widths formed as a result of the coupling between A and B excitons.
Therefore, the large splittings originated from the upper and lower branches appear, and the spectral shapes change with the film thickness.

  To determine the attribution of spectral peaks, we need to be careful of the radiative coupling between multicomponent excitons.
The mixing of A and B excitons is remarkable for the modes with large radiative corrections.
Therefore, the component density $D_{\sigma\lambda}$ becomes a significant index for a clear understanding of the relation between the excitonic system and their optical signals.
By utilizing $D_{\sigma\lambda}$, we can successfully demonstrate the determination of the attribution of the DFWM spectra, which would be one of the consideration elements for the existing discussions of the valence-band-ordering.

\vspace{3mm}
\section*{ACKNOWLEDGMENTS \label{6}}
  The authors thank Professor M. Nakayama, Professor M. Ashida, Professor M. Ichimiya, and Professor N. Yokoshi for their helpful discussions. 
This work was partially supported by a Grant-in-Aid for Scientific Research (A) No. 24244048 from Japan and by the Japan Society for Promotion of Science (JSPS).

\vspace{3mm}
\appendix*
\section{Reflectivity for Semi-Infinite System \label{7}}
  In this appendix, we derive the reflectivity for a semi-infinite system at normal incidence as demonstrated in Fig. \ref{reflectivity}(a) in Sec. \ref{3}.
The background electric field in the sample whose thickness is $d$ is written as
$
{\cal E}^{(0)}(z)={\cal E}_1e^{-iQ(z-d)}+{\cal E}_2e^{iQz},
$
where ${\cal E}_1$ (${\cal E}_2$) is an arbitrary constant that should be determined by the Maxwell's boundary conditions.
By solving Eq. (\ref{self-consistent}), $X_{\sigma\lambda}$ can be obtained as

\begin{eqnarray}
X_{A\lambda}
&
=
({\rm det}|\bm{S}|)^{-1}(E_{B\lambda}-\hbar\omega-i\Gamma_B)F^{(0)}_{A\lambda}, \\
X_{B\lambda}
&
=
({\rm det}|\bm{S}|)^{-1}(E_{A\lambda}-\hbar\omega-i\Gamma_A)F^{(0)}_{B\lambda}.
\end{eqnarray}
Then, we can rewrite Eq. (\ref{maxwell}) as

\begin{eqnarray}
{\cal E}(z,\omega)
&=&
{\cal E}^{(0)}(z,\omega)+
\nonumber
\\
&&
\hspace{-10mm}
\sum_{\lambda}
\{{\cal E}_1e^{iQd}J_\lambda(-Q)+{\cal E}_2J_\lambda(Q)\}g_{\sigma\lambda}(z),
\label{E}
\end{eqnarray}
where 

\begin{eqnarray}
J_\lambda(Q)
&=(-1+\frac{k^2_\lambda-Q^2}{\kappa^2_{\lambda}})\int g^\ast_{\sigma\lambda}(z)e^{iQz}dz,
\\
\kappa^2_{\lambda}
&=\frac{\hbar^4}{4M_AM_B}\frac{\Pi_{i=1,2,3}(k^2_\lambda-k^2_i)}{\Pi_{\sigma=A,B}(E_{\sigma\lambda}-\hbar\omega-i\Gamma_\sigma)}.
\end{eqnarray}
The summation over $\lambda$ can be converted to a contour integral in complex $k$-plane, where the contour picks up all the quantized values of $k$ on the real axis.
By deforming this contour to the one which picks up the poles of three polaritonic wavenumbers and $k=Q$, the integral can be rigorously evaluated \cite{Cho2,Ishihara2}.
Then the background electric field in the sample is canceled out, and Eq. (\ref{E}) is rewritten as

\begin{eqnarray}
{\cal E}(z,\omega)
&=&
\sum_{i}\zeta_{i}(\omega)\{{\cal E}_1(\sin{k_i}z+e^{iQd}\sin{k_i(d-z)})
\nonumber
\\
&&
+{\cal E}_2(e^{iQd}\sin{k_iz}+\sin{k_i(d-z)})\}, 
\end{eqnarray}
where

\begin{eqnarray}
\zeta_i(\omega)
&=
\frac{4M_AM_B}{\hbar^4}
\frac{\Pi_{\sigma=A,B}(E_{\sigma}+\frac{\hbar^2k^2_i}{2M_\sigma}-\hbar\omega-i\Gamma_\sigma)}{(k^2_i-k^2_{j(\neq i)})(k^2_i-k^2_{l(\neq i\neq j)})\sin{k_id}}.
\end{eqnarray}
Assuming that the background has a small absorption, i. e., $e^{iQd}\approx0$ ($d\rightarrow\infty$), ${\cal E}(z,\omega)$ can be written as

\begin{eqnarray} 
{\cal E}(z,\omega)
=
\sum_{i}\zeta_i(\omega)\{{\cal E}_1\sin{k_iz}+{\cal E}_2\sin{k_i(d-z)}\}.
\nonumber
\\ 
\end{eqnarray}
The Maxwell's boundary conditions of electromagnetic field at the incident surface provide the following relations by taking the limit of $d\rightarrow\infty$, and thus $e^{ik_id}\approx0$: 

\begin{eqnarray}
{\cal E}_{in}+{\cal E}_{re}=\alpha(\omega){\cal E}_2,
\hspace{5mm}
iq_{in}({\cal E}_{in}-{\cal E}_{re})\approx i\beta(\omega){\cal E}_2, 
\nonumber
\\
\label{mbc}
\end{eqnarray}
where
$
\alpha(\omega)
=
\sum_{i}\zeta_i(\omega),
$
and
$
\beta(\omega)
=
\sum_{i}k_i\zeta_i(\omega).
$
In Eq. (\ref{mbc}), ${\cal E}_{in}$ (${\cal E}_{re}$) is an arbitrary constant of the incident (reflectional) light, $q_{in}$ is a wavenumber of the incident light.
As a final expression, the reflectivity spectrum $R(\omega)$ at normal incidence can be obtained from
$
R(\omega)=|{\cal E}_{re}/{\cal E}_{in}|^2.
$


\end{document}